\documentclass[prl,showpacs,twocolumn]{revtex4-1}
\usepackage{amsmath,amsfonts,graphicx}
\usepackage[amssymb]{SIunits}

\newcommand{\rmd}{\,\mathrm{d}}

\begin{document}
\bibliographystyle{apsrev4-1}

\title{Hidden Lorentz symmetry of the Ho\v{r}ava - Lifshitz gravity}
\author{J.~Rembieli\'nski}\email{jaremb@uni.lodz.pl} 
\affiliation{Department of Theoretical Physics, University of Lodz,
Pomorska 149/153, 90-236 {\L}{\'o}d{\'z}, Poland}
\date{\today}

\begin{abstract}
In this letter it is shown that the Ho\v{r}ava-Lifshitz gravity theory admits Lorentz symmetry
preserving preferred global time foliation of the spacetime.
\end{abstract}
\maketitle

The possibility that gravity may exhibit a preferred foliation at its most fundamental level has
attracted a lot of attention recently, mainly due to the Ho\v{r}ava's papers \cite{raz,2,3} devoted to
gravity models characterized by certain specific anisotropic scaling between space and
time. The leading idea of the Ho\v{r}ava approach to the quantization of gravity is to achieve
power-counting renormalizability by modifying the graviton propagator. This is obtained by
adding to the action terms containing higher order spatial derivatives of the metric which, in
turn, naturally leads to the preferred co-dimension – one foliation $\mathcal{F}$ of
space-time manifold $\mathcal{M}$ topologically equivalent \cite{raz,2,3} to $R^1\times \Sigma$. The resulting theory,
known as the Ho\v{r}ava-Lifshitz (HL) gravity, is then invariant under a group of
diffeomorphisms $\mathcal{D}iff(\mathcal{F},\mathcal{M})$  preserving this foliation
\begin{equation}
\label{1}
\tilde{t}=\tilde{t}(t),\quad \tilde{x}^i=\tilde{x}^i(t,x^i)
\end{equation}
where $i=1,2,...,D$. 
The above mentioned anisotropic scaling characterizing HL gravity is of
the form
\begin{equation}
\label{2}
t\longrightarrow b^zt,\quad \boldsymbol{x}\longrightarrow b\boldsymbol{x}.
\end{equation}
Thus the (momentum) dimension $[t]=-z$, $[x^i]=-1$,
so the light velocity $c$ has the
dimension $[c]=z-1$. When $z$ equals the number of spatial dimensions $D$ the theory
becomes power-counting renormalizable provided all terms allowed are compatible with the
gauge symmetries in the action.

The HL theory is naturally described by the ADM  decomposition \cite{4} of the relativistic metric,
namely by the lapse function $N$ $([N]=0)$, the shift vector $N^i$ $([N^i]=[N_i]=z-1)$ and the
metrics $\gamma_{ij}$ $([\gamma_{ij}]=0)$ on the spacial slices $\Sigma$. In the HL gravity the lapse
$N=N(t)$ is only a
function of time $t$ which is constant along $\Sigma$ whereas the shift vector $N^i$
depends on the
spacetime point $(t,\boldsymbol{x})$. In terms of the ADM variables the metrics can be written as
\begin{multline}
\label{2a}
\rmd s^2=g_{\mu\nu}\rmd x^{\mu}\rmd x^{\nu}\\=-c^2N^2\rmd t^2+\gamma_{ij}(\rmd x^i+N^i\rmd t)(\rmd x^j+N^j\rmd t)
\end{multline}
The HL action, respecting the symmetries $\mathcal{D}iff(\mathcal{F},\mathcal{M})$ is \cite{raz,2,3}
\begin{equation}
\label{3}
S=\frac{2}{\kappa^2}\int \rmd t\rmd^D \boldsymbol{x}\sqrt{\gamma}N\left[\left(K_{ij}K^{ij}-\lambda K^2\right)-V\right],
\end{equation}
where $K=K_{\,\,i}^{i}$, $\lambda$ is a dimensionless coupling constant and 
\begin{equation}
\label{4}
K_{ij}=\frac{1}{2N}(\partial_t\gamma_{ij}-\nabla_iN_j-\nabla_jN_i)
\end{equation}
is the extrinsic curvature of the leaves hypersurface $\Sigma$. A scalar potential function $V$ is built
out of the spatial metrics, the spatial Riemann tensor and its covariant spatial derivatives
but is independent of the time derivatives of fields. For a review and extensions of the
Ho\v{r}ava's approach see \cite{5,6,7,8}. In the following we restrict ourselves to the physically
important $z=D=3$
case.

One of the problems of the Ho\v{r}ava-Lifshitz gravity is that this theory does not exhibit
Lorentz symmetry. A proposed way out of this situation is an appropriate preparation of the
potential to restore dynamically local Lorentz invariance in the low-energy limit \cite{raz,2,3}.
However, for each finite energy scale the Lorentz symmetry is in fact broken.
In this letter we suggest a way to overcome the difficulty with the Lorentz symmetry in the 
Ho\v{r}ava-Lifshitz gravity in a physically acceptable way. To do this
let us consider a coordinate independent solution
to the model defined by the action (\ref{3}) where the potential $V$ is chosen as in Ref.~\cite{raz} with the
cosmological constant equal to zero. Namely, let us choose the shift vector $\boldsymbol{N}$ as
\begin{equation}
\label{5}
\boldsymbol{N}=\frac{-c\,\boldsymbol{\epsilon}}{1-\boldsymbol{\epsilon}^2}
\end{equation}
with $0\leq \boldsymbol{\epsilon}^2<1$, while the lapse $N$ is given by
\begin{equation}
\label{6}
N=\frac{1}{\sqrt{1-\boldsymbol{\epsilon}^2}}.
\end{equation}
Furthermore, the space metrics is chosen as
\begin{equation}\label{7}
\gamma=(I-\boldsymbol{\epsilon}\otimes \boldsymbol{\epsilon}^{\rm{T}}),
\end{equation}
where T denotes transposition of the coordinate independent dimensionless column
vector $\boldsymbol{\epsilon}=(\epsilon^a)$, $a=1,2,3$. With help of the classical equations of motion \cite{9} it can be
verified that the equations (\ref{5}-\ref{7}) define the flat solution to the HL theory determined by (\ref{3}).
The spacetime metrics (\ref{2a}) takes the form
\begin{multline}
\label{8}
\rmd s^2=\zeta_{\alpha\beta}\rmd x^{\alpha}\rmd x^{\beta}\\=-c^2\rmd t^2-2c\boldsymbol{\epsilon}\cdot\rmd \boldsymbol{x}\rmd t+
\rmd \boldsymbol{x}^{\rm{T}}(I-\boldsymbol{\epsilon}\otimes \boldsymbol{\epsilon}^{\rm{T}})\rmd \boldsymbol{x}
\end{multline}
with the metric tensor
\begin{equation}
\label{9}
\zeta_{\alpha\beta}=\left(\begin{array}{c|c}
-1&-\boldsymbol{\epsilon}^{\rm{T}}\\\hline
-\boldsymbol{\epsilon}&I-\boldsymbol{\epsilon}\otimes \boldsymbol{\epsilon}^{\rm{T}}
\end{array}\right).
\end{equation}
Here $\alpha, \beta=0,1,2,3.$
It is easy to see that the metrics form (\ref{8}) is related to the Minkowski spacetime as well as
the space geometry is Euclidean. Now, let us consider the standard rotations
\begin{equation}
\label{10}
t'=t,\quad \boldsymbol{x}'=R\,\boldsymbol{x},\quad \boldsymbol{\epsilon}'=R\,\boldsymbol{\epsilon},
\end{equation}
where $R$ belongs to the group of orthogonal matrices, and the transfromations defined by 
\begin{subequations}
\begin{eqnarray}
t'&=&\frac{t}{a+\boldsymbol{a}\cdot\boldsymbol{\epsilon}},\label{11}\\
\boldsymbol{x}'&=&\left(I+\boldsymbol{a}\otimes\boldsymbol{\epsilon}^{\rm{T}}+
\frac{\boldsymbol{a}\otimes\boldsymbol{a}^{\rm{T}}}{1+a}\right)\boldsymbol{x}+\boldsymbol{a}ct,\label{12}\\
\boldsymbol{\epsilon}'&=&\frac{1}{a+\boldsymbol{a}\cdot\boldsymbol{\epsilon}}\left[\boldsymbol{\epsilon}+\boldsymbol{a}
\left(1+\frac{\boldsymbol{a}\cdot\boldsymbol{\epsilon}}{1+a}\right)\right],\label{13}
\end{eqnarray}
\end{subequations}
where $\boldsymbol{a}$ parametrizes the standard Lorentz boost $L(\boldsymbol{a})$
\begin{equation}
\label{14}
L(\boldsymbol{a})=\left(
\begin{array}{c|c}
a&\boldsymbol{a}^{\rm{T}}\\\hline {\tiny{}}&{\tiny{}}\vspace{-0.36cm}\\
\boldsymbol{a}&I+\frac{\boldsymbol{a}\otimes \boldsymbol{a}^{\rm{T}^{}}}{1+a}
\end{array}
\right),
\end{equation}
with $a=\sqrt{1+\boldsymbol{a}^2}$. 
It can be shown that the transformations (\ref{10}-\ref{13}) taken together form
the realization of the Lorentz group and it is obvious that they do not destroy the foliation $\mathcal{F}$.
Consequently, they do not affect absolute simultaneity and causality characteristic to the
Ho\v{r}ava-Lifshitz model. Moreover, the metrics (\ref{8}) is form invariant under the
transformations (\ref{10}-\ref{13}). We point out that in view of (\ref{10}-\ref{13}) the above transformations
form a nonlinear realization of the Lorentz group \cite{10,11}. Nonlinearity affects the
coordinate independent vector $\boldsymbol{\epsilon}$ only, whereas $\boldsymbol{x}$ and $t$ transform linearly. The nonlinear
realization (\ref{10}-\ref{13}) was firstly introduced in a different
context and form in \cite{12} and was applied to the localization
problem in the relativistic quantum mechanics \cite{21,22} as well as
to a Lorentz-covariant formulation of the statistical physics \cite{23}.
There is a simple relationship between the standard Lorentz transformations and those
given by (\ref{10}-\ref{13}). Indeed, introducing the new time coordinate by the affine
transformation (not belonging to the $\mathcal{D}iff(\mathcal{F},\mathcal{M})$)
\begin{equation}
\label{15}
t_E=t+\frac{\boldsymbol{\epsilon}\cdot\boldsymbol{x}}{c}
\end{equation}
we arrive at the standard Minkowski form of the metrics (\ref{8}). Moreover,
we can easily recover for $\boldsymbol{x}$ and $t_E$
the standard Lorentz transformations in the standard
pseudoorthogonal frame. Thus the time redefinition (\ref{15}) should be interpreted as the
change of distant clock synchronization \cite{13,14,15,16,17}. Consequently, the vector $\boldsymbol{\epsilon}$ plays the role
of the Reichenbach synchronization coefficient \cite{13,18}.

Now, it is not difficult to apply the above Lorentz covariant flat solution as the local
reference frame in a general case. This can be done by introducing the tetrad fields\
$\omega^{\alpha}=\lambda_{\,\,\mu}^{\alpha}\rmd x^{\mu}$ satisfying
\begin{equation}
\label{16}
\zeta_{\alpha\beta}\omega^{\alpha}\omega^{\beta}=g_{\mu\nu}\rmd x^{\mu}\rmd x^{\nu},
\end{equation}
with $\zeta_{\alpha\beta}$
and $g_{\mu\nu}$
given by (\ref{8}) and (\ref{2a}) respectively. The solution has the form
\begin{subequations}
\begin{eqnarray}
\label{17}
\omega^0&=&(cN-\epsilon^ae^{a}_{i}N^i)\rmd t-\epsilon^ae^{a}_{i}\rmd x^i,\\ \omega^a&=&e^a_i(\rmd x^i+N^i\rmd t),
\end{eqnarray}
\end{subequations}
where the triads $e^a_i$
determine the space metrics
$e^a_ie^a_j=\gamma_{ij}$. The tetrads $\omega^{\alpha}$
transform with respect to the index $\alpha$ according to the law (\ref{10}-\ref{13}) treated as the frame
transformations. Notice, that in general the synchronization vector $\boldsymbol{\epsilon}$ is frame dependent
because it transforms from frame to frame according to the formula (\ref{13}). In particular, we
can specify the boost parameter $\boldsymbol{a}$ to obtain the synchronization vector $\boldsymbol{\epsilon}$ equal to zero in a
distinguished frame. In this peculiar frame related to the preferred foliation $\mathcal{F}$ the
Einstein synchronization convention applies. It can be shown
\cite{21,22} that the synchronization vector $\boldsymbol{\epsilon}$
can be related to the velocity of the preferred frame. Finally, let us stress that the synchronization
change (\ref{15}) does not affect the physical content of theory on the classical level because of
the conventionality of the synchronization procedure \cite{13,14,15,16,17,18}. However, it breaks the
quantization procedure essential to the Ho\v{r}ava approach. This can indicate that result of
quantization depends on the adapted synchronization scheme.
Concluding, the Ho\v{r}ava-Lifshitz gravity admits Lorentz symmetry preserving preferred
global time foliation of the spacetime. This symmetry can be related to the standard Lorentz
transformations by the frame – dependent change of synchronization (\ref{15}) to the Einstein
one. However, (\ref{15}) breaks the preferred foliation of the HL gravity. Thus the HL theory
forces Lorentz symmetry realized in the synchronization scheme related to the
transformation laws (\ref{10}-\ref{13}). Our observation can be also applied to the causal dynamical
triangulation theory \cite{19}, where the global time foliation is assumed too (however see \cite{20}).

\begin{acknowledgments}
The author is grateful to Bogus{\l}aw Broda and Krzysztof Kowalski for discussion and to Jerzy Jurkiewicz for helpful remarks concerning the causal dynamical triangulation theory.
\end{acknowledgments}


%

\end{document}